\documentclass[aps,prl,twocolumn,superscriptaddress]{revtex4-1}

\usepackage{graphicx}
\usepackage{amsmath}
\usepackage{amssymb}
\usepackage{braket}
\usepackage{subfigure}
\usepackage{hyperref}

\begin{document}

\title{Visualizing Neural Network Developing Perturbation Theory}
\author{Yadong Wu}
\affiliation{Institute for Advanced Study, Tsinghua University, Beijing, 100084, China}

\author{Pengfei Zhang}
\affiliation{Institute for Advanced Study, Tsinghua University, Beijing, 100084, China}

\author{Huitao Shen}
\affiliation{Department of Physics, Massachusetts Institute of Technology, Cambridge, Massachusetts 02139, USA}

\author{Hui Zhai}
\affiliation{Institute for Advanced Study, Tsinghua University, Beijing, 100084, China}
\affiliation{Collaborative Innovation Center of Quantum Matter, Beijing, 100084, China}

\date{\today}

\begin{abstract}
Motivated by the question that whether the empirical fitting of data by neural networks can yield the same structure of physical laws, we apply neural networks to a quantum mechanical two-body scattering problem with short-range potentials---a problem by itself plays an important role in many branches of physics. After training, the neural network can accurately predict $ s $-wave scattering length, which governs the low-energy scattering physics. By visualizing the neural network, we show that it develops perturbation theory order by order when the potential depth increases, without solving the Schr\"odinger equation or obtaining the wavefunction explicitly. The result provides an important benchmark to the machine-assisted physics research or even automated machine learning physics laws. 
\end{abstract}

\maketitle

Human physicists have made great achievements in discovering laws of physics during the last several centuries. Based on experimental observations, they interpret data and build theories using logical inference and rigorous mathematical reasoning. Artificial intelligence (or machine learning), which has been extensively applied to physics researches recently \cite{classify1,classify2,classify3,classify4,classify5,classify6,classify7,classify8,classify9,classify10,classify11,classify12,classify13,classify14,classify15,classify16,classify17,classify18,classify19,classify20,accelerate1,accelerate2,accelerate3,accelerate4,accelerate5,accelerate6,numerics4,numerics1,numerics2,numerics3,new1,new2,new3,new4,new5,new6,new7}, interprets data empirically by, for example, fitting on statistical models.  

While physicists hope that the artificial intelligence could help discover new physics in unexplored areas, there are certainly challenges to overcome due to the different nature of ``knowledge'' among human physicists and artificial intelligence. First, the great complexity of machine learning models, such as deep neural networks with more than millions of fitting parameters, makes it difficult for humans to understand whether machines have successfully captured the underlying laws of physics. Second, despite the great success of artificial intelligence in many disciplines outside physics \cite{MachineNature}, its weakness \cite{weak1,weak2} leads people to believe that it lacks the ability of sophisticated reasoning \cite{science}. Therefore, it is important to ask the question that whether these two approaches of distilling knowledge from data will yield the same answer when facing the same problem. The positive answer will mark an important step forward to the machine-assisted physics research or even automated machine learning physics laws.

Motivated by this question, we let the machine supervisedly learn quantum mechanical scattering problem. We show that the machine could not only correctly produce the scattering length for a given interaction potential, but also give a human tractable decision-making process. More concretely, the machine will develop perturbation theory order by order with increasing the potential strength.

\textit{Two-body Scattering Problem. }
The scattering problem plays an important role in many branches of physics, including atomic and molecular physics, nuclear physics and particle physics \cite{Braaten}. In this work, we focus on the two-body scattering problems in three dimensions with spherical potentials. Consider the Schr\"odinger equation in the relative frame
\begin{equation}
\left(-\frac{\hbar^2}{2\bar{m}}\nabla^2+V(r)\right)\Psi=E\Psi,
\end{equation}
where $\bar{m}$ is the reduced mass of two particles and $V(r)$ is the interaction potential. The wavefunction can be expanded through partial wave decomposition \cite{Landau} as
\begin{equation}
\Psi({\bf r})=\sum\limits_{l=0}^{+\infty}\frac{\chi_{kl}(r)}{kr}\mathcal{P}_l(\cos\theta),
\end{equation}
where $E=\hbar^2 k^2/(2\bar{m})$, $\chi_{kl}$ is the radial wavefunction and $\mathcal{P}_l(\cos\theta)$ represents the angular part for each partial wave. For $s$-wave scattering, one can show that the asymptotic behavior of the radial wavefunction $\chi_{k,l=0}\propto \sin(kr+\delta_k)$. That is to say, $\delta_k$ is the only quantity that fixes the scattering wavefunction at low-energy given potential $V(r)$. We can therefore determine the important quantity of the $s$-wave scattering length $a_\text{s}$ as $\tan\delta_k=-ka_\text{s}+o(k^2)$ \cite{Landau}.

In this way, we establish a mapping between $V(r)$ and $a_\text{s}$, which is to be learned by the neural network. In the following, we first train neural networks with $V(r)$ as the input and $a_\text{s}$ as the output. Here
$a_\text{s}$ is obtained by numerically solving the Schr\"odinger equation \cite{SM}. In practice, $a_\text{s}$ can be extracted from experimental data of low-energy collision. After training, the neural network can predict $a_\text{s}$ directly from $V(r)$. 

\textit{Perturbation Theory.} Before presenting results of neural networks, let us first review a textbook method for computing the scattering length. Aside from numerically solving the Schr\"odinger equation, one can also compute the scattering amplitude $\mathcal{T}$ analytically through perturbative Born expansion \cite{Landau}. In fact, the perturbation theory is by far the most well-established approach to compute observables in interacting quantum mechanics or quantum field theory. Particularly in our two-body scattering problem, $\mathcal{T}$ is given by  
$\mathcal{T}=V+VG_0V+VG_0VG_0V+\dots,$
where $G_0$ is the Green's function for free Hamiltonian with $V(r)=0$. Keeping only the $s$-wave component and taking the low-energy limit, we obtain $a_\text{s}=2\pi^2\langle 0|\mathcal{T}|0\rangle$, where $|0\rangle$ denotes the zero-momentum state. Here and throughout the rest of this work, we take $m=\hbar=1$ for simplicity. 
Combining these equations, we get $a_\text{s}=a^{(1)}_\text{s}+a^{(2)}_\text{s}+\dots$, with 
\begin{equation}
a^{(1)}_\text{s}=2\pi^2\langle 0|V|0\rangle=\int V(r) r^2 dr,
\label{first_Born_Eq}
\end{equation}
as the first order Born approximation and 
\begin{align}
a^{(2)}_\text{s}&=2\pi^2\langle 0|VG_0V|0\rangle\nonumber\\
&=-\frac{1}{2}\int V(r)V(r^\prime)\mathcal{K}(r,r^\prime)drdr^\prime, \label{second_Born}
\end{align}
as the second order Born approximation. Here 
\begin{equation}
\mathcal{K}(r,r^\prime)=rr^\prime(r+r^\prime-|r-r^\prime|).
\end{equation}
The Born expansion is based on the potential strength, and is a very good approximation for weak potentials.

\begin{figure}[t]
\includegraphics[width=0.48\textwidth]{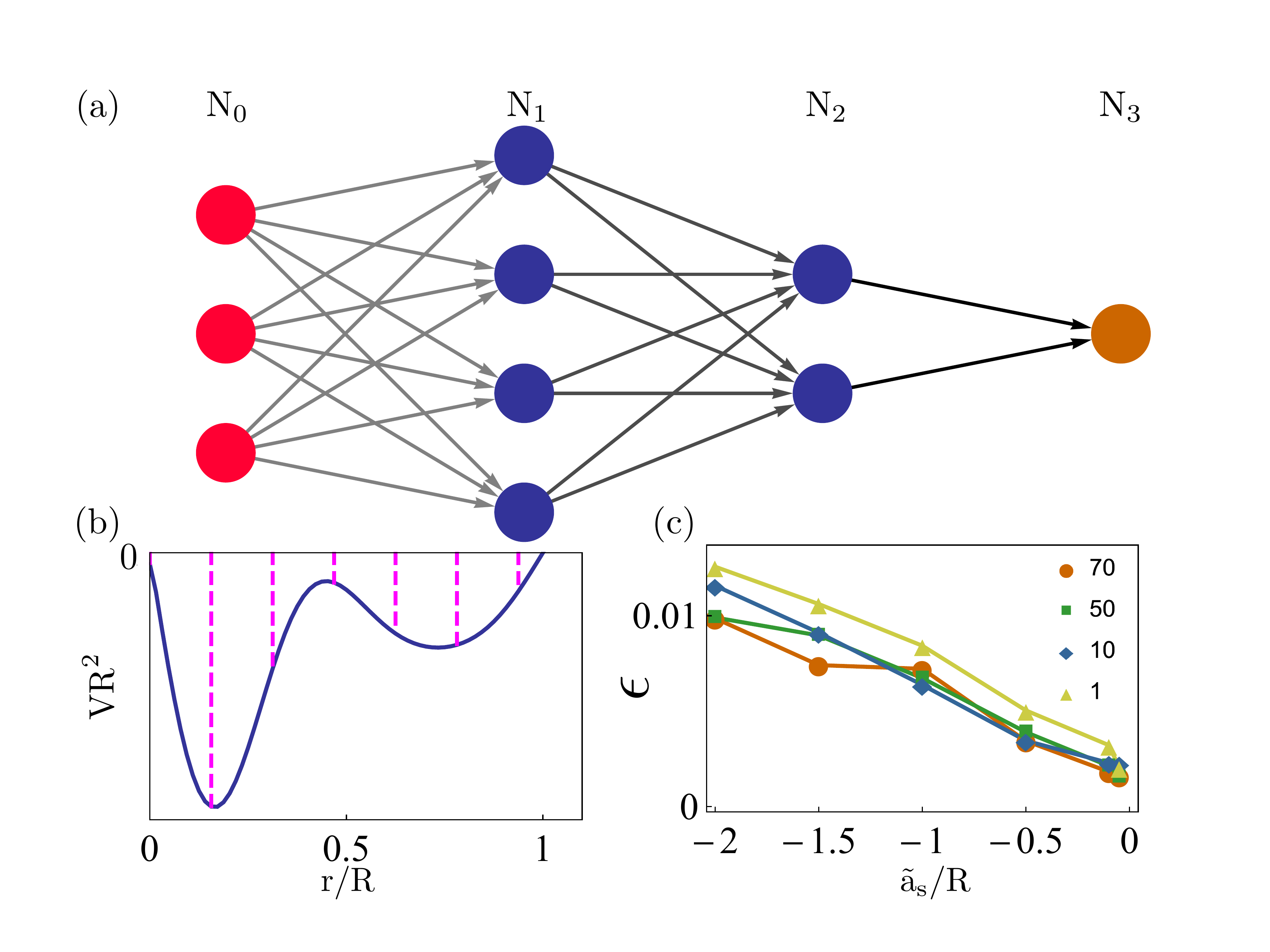}
\caption{(a) Schematic of the fully-connected neural network. (b) Schematic of the input data, which is a finite range potential discretized at $N_0$ different points. (c) The error $\epsilon$ of the neural network after training as a function of maximum scattering length $\tilde a_\text s$, with different number of neurons $N_1$ in the first hidden layer. See main text for the definition of $\epsilon$ and $\tilde a_\text s$.}
\label{network}
\end{figure}

\textit{Neural Network.} In order to keep the size of the input data finite, we consider short-range attraction potentials $V(r)$, i.e. $V(r)<0$ for $r<R$ and $V(r)=0$ for $r\geq R$. For simplicity, we restrict the depth of $V(r)$ to be before the first resonance so that $a_{\text s}<0$. We first generate the potential randomly \cite{SM} and then discretize it uniformly along $r=0$ to $r=R$ into $N_0$ pieces, as shown in Fig.~\ref{network}(b). The input data is thus an $N_0$-dimensional vector
\begin{equation}
{\bf V}=\left(V\left(\frac{R}{N_0}\right),V\left(\frac{2R}{N_0}\right),\dots,V\left(R\right)\right)^\text{T}.
\end{equation}
The potential depth is characterized by its mean value as $V_\text{c}=-(1/N_0)\sum_{i=1}^{N_0}V_i>0$, where $V_i$ is the $i$-th component of ${\bf V}$. Before the first resonance, the larger $V_\text{c}$, the larger the absolute value of scattering lengths in general.

We use a fully-connected neural network as our machine learning model, whose structure is schematically shown in Fig.~\ref{network}(a). It is composed of three fully-connected layers. The effect of the first (hidden) layer can be mathematically summarized as ${\bf Y}=f(W{\bf V}+{\bf B}),$ where $ {\bf V} $ is the $ N_0 $-dimensional input vector defined previously, ${\bf Y}$ is an $N_1$-dimensional output vector, $W$ is an $N_1\times N_0$ weight matrix and ${\bf B}$ is an $N_1$-dimensional bias vector. $ f $ is the activation function that is applied element-wisely to the vector and is chosen to be rectified linear units $f(x)=\text{max}\{0,x\}$. A similar hidden layer taking $ {\bf Y} $ as its input comes after the first hidden layer. Its weight matrix is of size $ N_2\times N_1 $, and its output is linearly mapped to the final single value output, which is interpreted as the scattering length $a_\text{s}$. In our work, we set $N_0=64, N_2=32$ fixed and change $N_1$ of the neural network.

We train the neural network by minimizing the mean squared error on $ 3\times10^4 $ randomly generated $ \{{\bf V}, a_\text s^{\text{T}}(\mathbf{V})\} $ pairs in units of $\{1/R^2,R\}$, where $ a_\text s^{\text{T}}(\mathbf{V}) $ is the scattering length obtained from solving the Schr\"odinger equation given discretized potential $\mathbf{V}$. In the training data set, $ a_\text s^{\text{T}}(\mathbf{V}) $ is uniformly distributed in the range of $(\tilde{a}_\text s,0)$. After training, the neural network should be able to make a prediction of the scattering length, denoted as $a_\text s^{\text{P}}(\mathbf{V})$, directly from the input potential $\mathbf{V}$. This prediction may deviate from the true scattering length $a_\text s^{\text{T}}(\mathbf{V})$. We characterize the performance of the trained neural network by calculating the averaged error $\epsilon$ defined as
\begin{align}
\epsilon=\frac{1}{N_\text{t}}\sum_l\left| \frac{a_\text s^\text{P}(\mathbf{V}^l)-a_\text s^\text{T}(\mathbf{V}^l)}{a_\text s^\text{T}(\mathbf{V}^l)}\right |,
\end{align}
on a test set of similar $ \{{\bf V}, a_\text s^{\text{T}}(\mathbf{V})\} $ pairs. Here ${\bf V}^l$ denotes the $l$-th potential in the test set and $N_\text{t}=10^3$ is the size of the testing set. As shown in Fig. \ref{network}(c), the network can always predict the scattering length $a_s$ with high accuracy for different $\tilde{a}_\text s$ and with different number of neurons in the first layer (denoted by $N_1$).

\begin{figure}[t]
\includegraphics[width=0.45\textwidth]{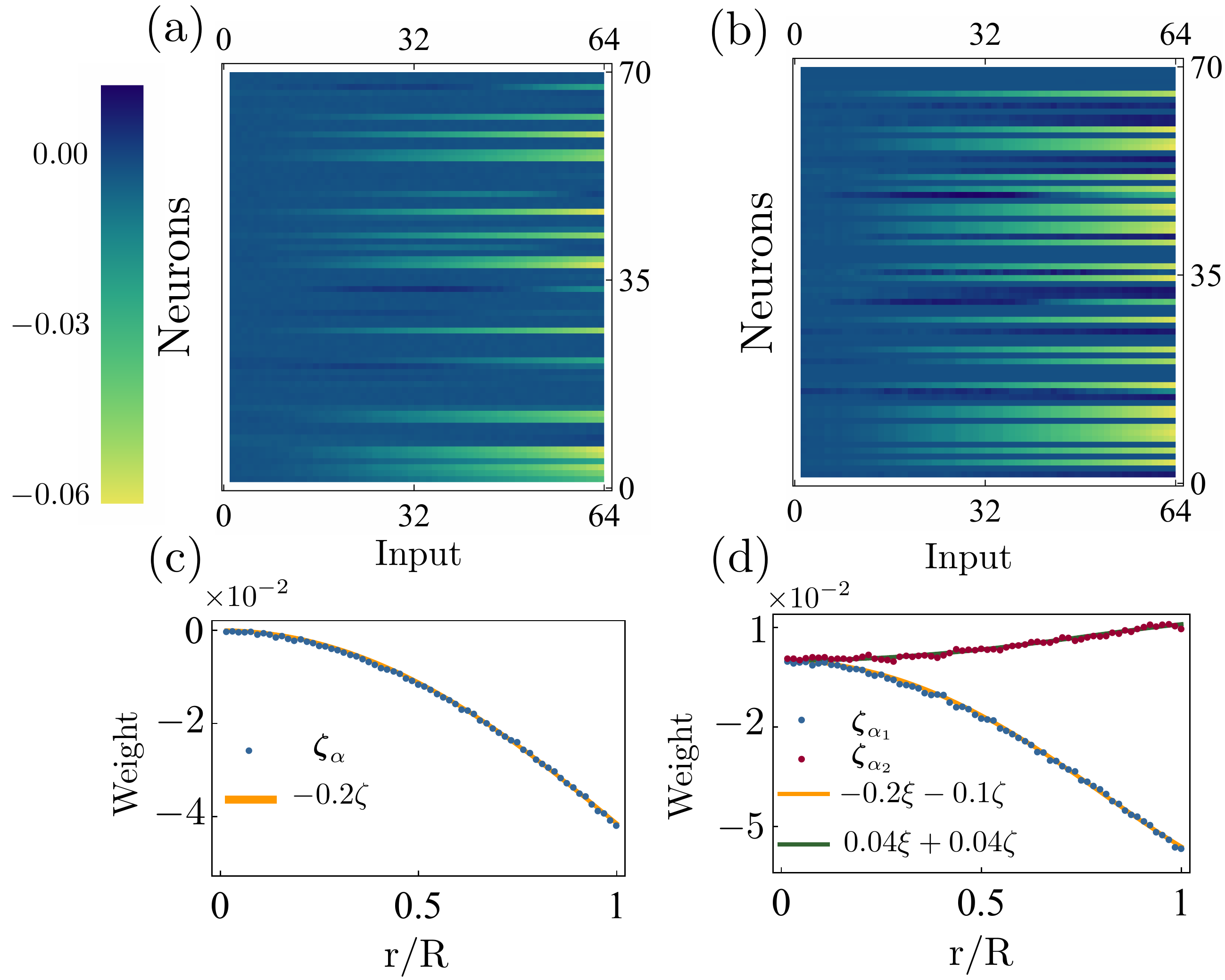}
\caption{Visualization of the matrix $W$ ((a) and (b)) and a typical $N_0$-dimensional vector $\boldsymbol{\zeta}_\alpha=\{W_{\alpha i},i=1,\dots,N_0\}$ as a function of $r/R=i/N_0$ ((c) and (d)), for $\tilde{a}_\text{s}/R=-0.1$ ((a) and (c)) and for $\tilde{a}_\text{s}/R=-0.5$ ((b) and (d)). The solid line in (c) is a fitting of $\boldsymbol{\zeta}_{\alpha=46}$. The yellow and green solid lines in (d) are fittings of $\boldsymbol{\zeta}_{\alpha_1=46}$ and $\boldsymbol{\zeta}_{\alpha_2=54}$ as linear combinations of $\boldsymbol{\xi}$ and $\boldsymbol{\zeta}$. 
}
\label{mapping}
\end{figure}

To understand how the neural network computes the scattering length, it is very informative to visualize the weight matrix $W$ in the first hidden layer that maps the input data ${\bf V}$ to the first intermediate vector ${\bf Y}$. We plot $W$ in Fig.~\ref{mapping} for two neural networks trained on a very small magnitude of $ \tilde{a}_\text{s}/R=-0.1 $ and on a relatively large magnitude of $ \tilde{a}_\text{s}/R=-0.5$, respectively. For both networks, most of their matrix elements automatically vanish and are thus not responsible for producing the scattering length. In the following, we analyze the remaining matrix elements for these two scenarios. 

\textit{First Order Born Approximation.} Let us first focus on the case of shallow potentials. For convenience, we represent $ W $ using $ N_0 $-dimensional row vectors, known as ``neurons'': $ W=(\boldsymbol{\zeta}_1, \boldsymbol{\zeta}_2, \ldots, \boldsymbol{\zeta}_{N_1})^T $. The effect of the first (hidden) layer can then be represented as $Y_i = f ( \boldsymbol{\zeta}_i ^T\mathbf V + B_i)$. The $j$-th component of the $i$-th neuron is denoted as $ \zeta_{i,j} $. From Fig.~\ref{mapping}(a), one can see that nearly all non-vanishing $\boldsymbol{\zeta}_\alpha$ behave similarly. A typical $\boldsymbol{\zeta}_\alpha$ is plotted in Fig.~\ref{mapping}(c). 

Importantly, we find that typical $\zeta_{\alpha,j}$ can be quite accurately described by $\sim j^2$. As shown in Fig.~\ref{mapping}(c), $\boldsymbol{\zeta}_\alpha$ and normalized $ \boldsymbol{\zeta}\propto (1,4,...,N_0^2)$ correspond very well. This means that the $\alpha$-th neuron in the first hidden layer performs the calculation
\begin{equation}
Y_\alpha=\sum\limits_{j}W_{\alpha j}V_j\propto \sum\limits_j j^2 V\left(\frac{j R}{N_0}\right).
\end{equation}
This is precisely the discrete version of the first order Born approximation shown in Eq.~\eqref{first_Born_Eq}. If one uses similar neural network to study scattering volume or super-volume for $p$-wave or $d$-wave scattering, one obtains that typical ${\boldsymbol \zeta}_{\alpha}$ for shallow potential behaves as $\sim j^4$ or $\sim j^6$, respectively, which are both consistent with the first order Born approximation \cite{SM}. 

\begin{figure}[t]
\includegraphics[width=0.48\textwidth]{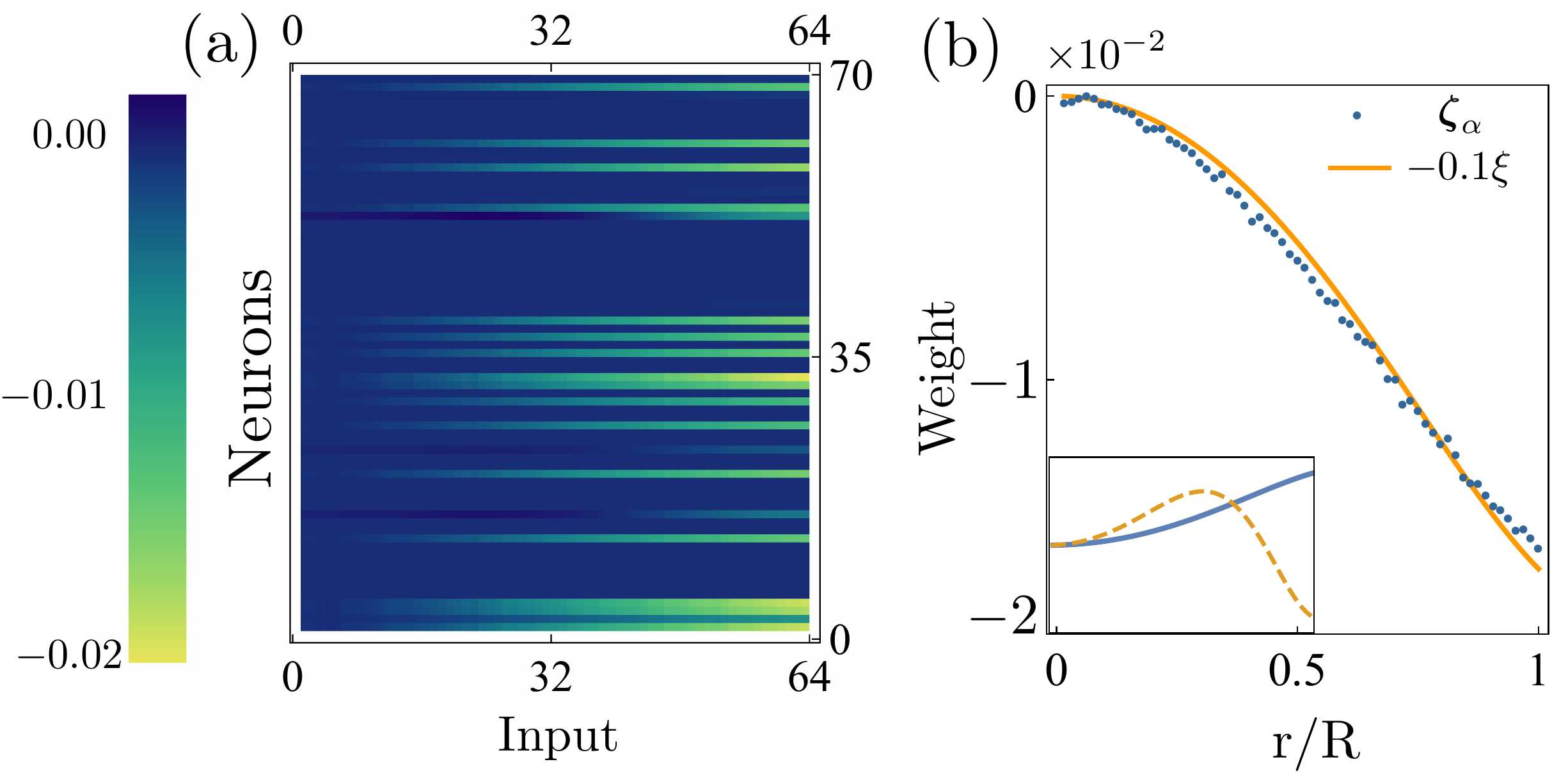}
\caption{Neural network with the same architecture as that in Fig. \ref{mapping}, but trained with $a_\text s^{(2)}$ as the output instead of $a_\text{s}$. (a) Visualization of the matrix $W$; (b) A typical vector $\boldsymbol{\zeta}_{\alpha=32}$ as a function of $r/R=i/N_0$ for $\tilde{a}_\text{s}/R=-0.5$. It is compared with the eigenvector of $K$-matrix with the largest eigenvalue $\sim \boldsymbol{\xi}$ shown by the solid line.  The inset of (b) are the eigenvectors associated with first two largest eigenvalues. For $N_0=64$, the blue curve shows $\sim \boldsymbol{\xi}_1$ with the largest eigenvalue $\Lambda_1=29.5$, and the yellow dashed curve shows $\sim \boldsymbol{\xi}_2$ with the second largest eigenvalue $\Lambda_2=1.8$.  }
\label{second}
\end{figure}

\textit{Second Order Born Approximation.} As the potential depth increases, more features begin to emerge in $W$. There are many neurons whose behavior cannot be fitted by $j^2$ discussed in the previous section, shown in Fig. \ref{mapping} (b) and (d). It is natural to suspect that this new neuron pattern comes from the second order Born approximation. To verify this, we first notice that the discretized version of Eq.~\eqref{second_Born} is
\begin{equation}
a^{(2)}_\text{s}=-\sum\limits_{ij}V\left(\frac{iR}{N_0}\right)V\left(\frac{jR}{N_0}\right)K_{ij},
\label{disB2}
\end{equation}
where $K$ is a $N_0\times N_0$ matrix constructed as
\begin{equation}
K_{ij}=\frac{1}{N^3_0}ij(i+j-|i-j|).
\end{equation} 
$K$ matrix can be diagonalized, and let us denote $\boldsymbol{\xi}_\alpha$ as its eigenvector associated with eigenvalue $\Lambda_\alpha$. In this way, we can rewrite Eq.~\eqref{disB2} as
\begin{equation}
a^{(2)}_\text{s}=-{\bf V}^T K {\bf V}=-\sum\limits_{\alpha}\Lambda_\alpha (\boldsymbol{\xi}^T_\alpha {\bf V})^2.
\label{disB22}
\end{equation}
Moreover, since the largest eigenvalue $\Lambda_1$ is an order of magnitude larger than the second largest eigenvalue $\Lambda_2$, i.e. $\Lambda_1/\Lambda_2\sim16.38$, we can approximate Eq.~\eqref{disB22} further by only keeping the contribution of the largest eigenvalue as
\begin{equation}
a^{(2)}_\text{s}\propto (\boldsymbol{\xi}_1^T {\bf V})^2. 
\label{Simple2}
\end{equation}

Indeed, if we train the neural network with the same architecture as before, but using $a^{(2)}_\text{s}$ calculated using Eq.~\eqref{second_Born} as the output instead of the exact $a_\text{s}$, the neural network could compute $a^{(2)}_\text{s}$ accurately. In this new network, neurons in the first hidden layer have only one pattern, and can be well fitted by $ \boldsymbol{\xi}_1 $ defined above, as illustrated in Fig. \ref{second}. In this way, neurons in the first hidden layer computes  $\boldsymbol{\xi}^T{\bf V}$ with $\boldsymbol{\xi}\equiv \boldsymbol{\xi}_1$ and the second hidden layer simply fits a function $x^2$ in order to compute $a^{(2)}_\text{s}$ in the second order Born approximation. 

Equipped with the vector $\boldsymbol{\zeta}$ introduced for the first order Born approximation and the vector $\boldsymbol{\xi}$ introduced for the second order Born approximation, we find neurons in $W$ with new features for a deeper potential can be well fitted by $ \alpha \boldsymbol{\zeta}-b\boldsymbol{\xi} $, as shown in Fig.~\ref{mapping}(d). In this way, in the first hidden layer, all ingredients for the first order and the second order Born approximation are already computed. With a proper polynomial function implemented in the second hidden layer, the neural network produces $a_\text{s}=a^{(1)}_\text{s}+a^{(2)}_\text{s}$ as its final output.  

The key observations so far can be summarized as follows: 
(i) When the potential is shallow, the neural network only captures the first order perturbation as it is already accurate enough to reach certain level of accuracy. (ii) As the potential becomes deeper, the neural network gradually develops the structure to capture at least the second order perturbation. It is quite remarkable that, by empirical fitting, the neural network develops the same kind of perturbation theory as that developed by human physicists. At least in this example, we understand how the neural network works and provides a positive answer to the question raised in the introduction, namely, human physicists and neural networks can yield the same answer when facing the same problem.  

\begin{figure}[t]
\includegraphics[width=0.48\textwidth]{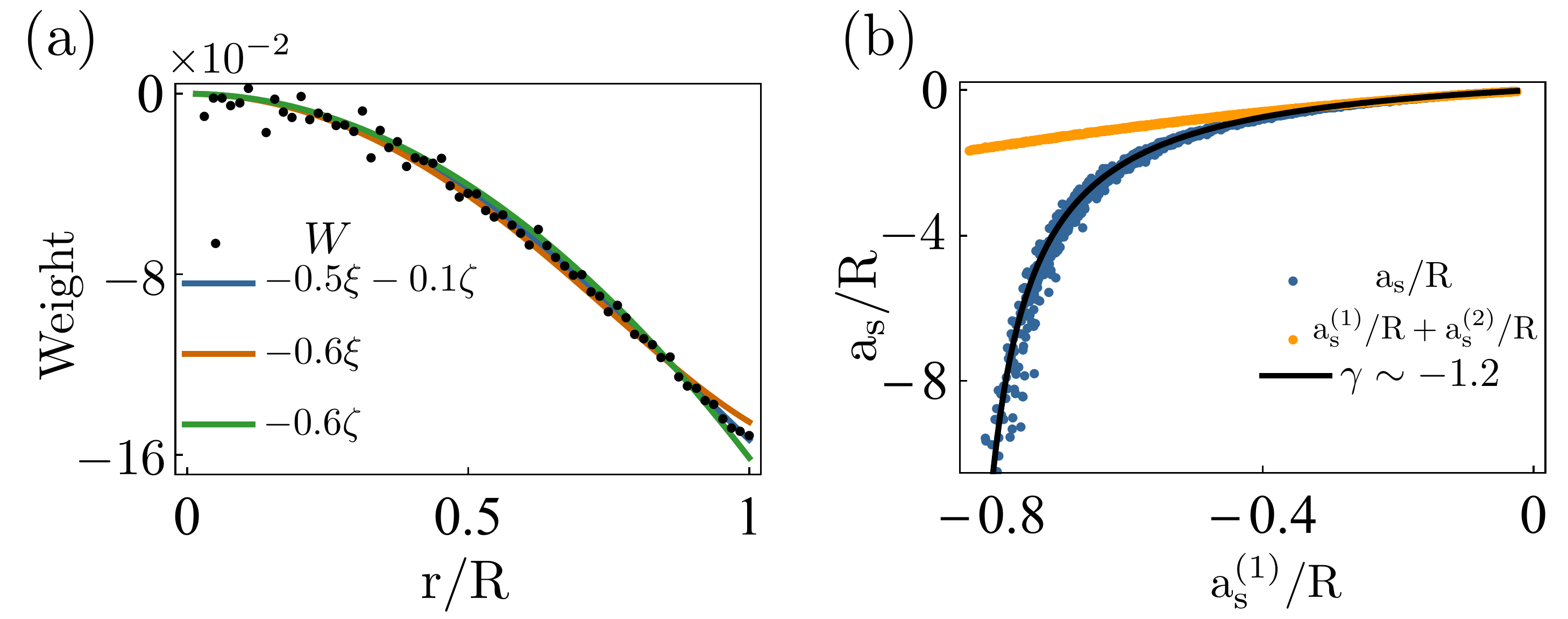}
\caption{(a) When the first layer only contains one neuron, $W$ becomes a $N_0$-dimensional vector. Here we compare $W$ and different linear combinations of $\boldsymbol{\zeta}$ and $\boldsymbol{\xi}$. (b) The relation between the actual scattering length $a_\text{s}$ and $a^{(1)}_\text{s}$ from the first order Born approximation. The black line is the empirical formula Eq. \eqref{Fempiral} and the yellow dots are scattering lengths computed from first two orders of Born approximation.}
\label{empiral}
\end{figure}

\textit{An Empirical Formula.} Last but not the least, we bring out a different aspect: If one can tolerant certain amount of error, the neural network can also inspire the empirical formula that works quite well even beyond the perturbative regime, although the formula lacks rigorous mathematical reasoning. 

Here the formula is inspired by the fact that even if we reduce the first hidden layer to have only one neuron, the neural network can still predict scattering lengths quite accurately with errors only increasing by a little amount (Fig.~\ref{network}(c)). In this extreme case, the effect of the hidden layer is simply a vector inner product. In Fig.~\ref{empiral}(a), we compare $W$ with $\boldsymbol{\zeta}$ and $\boldsymbol{\xi}$. After proper scaling, they actually behave similarly. This implies that they can replace with each other interchangeably without introducing much error. Therefore, (i) if we replace $W$ with $\boldsymbol{\zeta}$, this simple neural network produces nothing but $a^{(1)}_\text{s}$, and (ii) if we replace $\boldsymbol{\xi}$ also with $\boldsymbol{\zeta}$, by Eq.~\eqref{Simple2} $ a^{(2)}_\text{s} $ can be represented by $a^{(1)}_\text{s}$ as $a^{(2)}_\text{s}\propto (a^{(1)}_\text{s})^2/R$. With these we have 
\begin{align}
a_\text{s}&=a^{(1)}_\text{s}+\gamma (a^{(1)}_\text{s})^2/R+O((a^{(1)}_\text{s}/R)^3)\nonumber\\
&\approx \frac{a^{(1)}_\text{s}}{1-\gamma a^{(1)}_\text{s}/R}, \label{Fempiral}
\end{align}
where $\gamma$ is a fitting parameter. 

Equation \eqref{Fempiral} is our empirical formula for scattering lengths. It takes $ a^{(1)}_\text{s} $ as the input and produce $ a_\text{s} $ as the output. To verify this formula, we first generate many interaction potentials randomly, then compute the corresponding $a_\text{s}$ exactly from solving the Schr\"odinger equation, and $a^{(1)}_\text{s}$ from Eq.~\eqref{first_Born_Eq}. If Eq.~\eqref{Fempiral} is a good approximation, these data will collapse into a single curve that can be fitted by Eq.~\eqref{Fempiral}. This is indeed the case as is shown in Fig.~\ref{empiral}(b), where the fitting parameter is taken to be $\gamma=-1.2$. Interestingly, Equation \eqref{Fempiral} works quite well even when $a_\text{s}$ is an order of magnitude larger than $a^{(1)}_\text{s}$, that is to say, when the potential is deep enough to be beyond the weakly interacting regime. To compare, we also plot $\{a^{(1)}_\text{s}+a^{(2)}_\text{s},a^{(1)}_\text{s}\}$ from the same set of potentials in Fig.~\ref{empiral}(b). Clearly, the validity of the empirical formula Eq.~\eqref{Fempiral} goes significantly beyond the second order Born approximation.

\textit{Outlook.} In summary, we have trained a neural network to predict the $s$-wave scattering length directly from the interaction potential. By visualizing its weight matrix, we demonstrate that the neural network develops perturbation theory order by order as the interaction strength increases. Moreover, the performance of the neural network also inspires us to derive a simple approximate formula for scattering lengths that works well beyond the perturbative regime. 

Guaranteed by the great expressibility of neural networks \cite{anyfunc1,anyfunc2}, our formalism can be generalized to study more complicated quantum few-body and many-body problems. The former is important in atomic and molecular physics, as well as quantum chemistry. The later is connected to machine learning phases of matter, which has already become a topic of intensive interests \cite{classify1,classify2,classify3,classify4,classify5,classify6,classify7,classify8,classify9,classify10,classify11,classify12,classify13,classify14,classify15,classify16,classify17,classify18,classify19,classify20}. Our understanding of how neural network works in this simple problem can shed light on understanding more sophisticated problems better.

\textit{Acknowledgment.} We thank Ning Sun, Ce Wang and Jinmin Yi for helpful discussions. This work is supported by MOST under Grant No. 2016YFA0301600 and NSFC Grant No. 11734010. H.S. thanks IASTU for hosting his visit to Beijing, where the key parts of this work were done. 


\end{document}